\documentstyle[12pt,twoside]{article}
\begin{document}
\begin{center}
{\Large\bf 

QUANTUM COSMOLOGICAL PERFECT\\[5PT]

FLUID MODELS\\[5PT]}

\medskip

{\bf F.G. Alvarenga$^\dag$\footnote{e-mail: flavio@cce.ufes.br},
J. C. Fabris$^\dag$\footnote{e-mail: fabris@cce.ufes.br}, N.A.
Lemos$^\ddag$\footnote{e-mail: nivaldo@if.uff.br} and G. A. 
Monerat$^\ddag$\footnote{e-mail: monerat@if.uff.br}}  \medskip

\dag Departamento de F\'{\i}sica, Universidade Federal do Esp\'{\i}rito Santo, 
CEP29060-900, Vit\'oria, Esp\'{\i}rito Santo, Brazil \medskip\\

\ddag Departamento de F\'{\i}sica, Universidade Federal Fluminense,
CEP24210-340, Niter\'oi, Rio de Janeiro, Brazil

\medskip

\end{center}

\begin{abstract}

Perfect fluid Friedmann-Robertson-Walker quantum cosmological models for an arbitrary barotropic equation of
state $p = \alpha\rho$ are constructed using  Schutz's variational formalism.
In this approach the notion of time can be recovered. By superposition of stationary states, finite-norm wave-packet solutions to the Wheeler-DeWitt equation are found. The behaviour
of the scale factor is studied by applying the many-worlds and the
ontological interpretations of quantum mechanics. Singularity-free models are obtained
for $\alpha < 1$. Accelerated expansion at present requires $- 1/3 > \alpha > - 1$.

\vspace{0.7cm}

PACS number(s): 04.20.Cv., 04.20.Me

\end{abstract}

\section{Introduction}

It is generally believed that the singularity which plagues the
standard cosmological scenario may be avoided by taking into account
quantum effects when, going backward in time, the Universe reaches the Planck scale.
Since there is no consistent quantum theory of gravity until now,
this possibility
remains a speculation. However,  important information
concerning the singularity problem is expected to be obtained through the quantum cosmological approach \cite{halliwell}. In quantum
cosmology the Hamiltonian formulation of general relativity is
employed through the ADM decomposition of the geometry \cite{adm}, and a Schr\"odinger-like equation, the Wheeler-DeWitt
equation, is constructed which determines the wave function of the
Universe as a whole.

\par
However, quantum cosmology suffers from many technical and
conceptual problems. Technically, the Wheeler-DeWitt equation of quantum gravity  is
a functional differential equation defined in the so-called superspace, the space
of all possible three dimensional spatial metrics, and no general solution
in this superspace is known so far. In quantum cosmology this problem is circumvented
by freezing out an infinite number of degrees of freedom by symmetry requirements, leaving only a few
ones to be submitted to the quantization process. This procedure defines the minisuperspace, where  exact solutions
can often be determined. On the other hand, upon applying the ADM decomposition
general covariance is lost, and in most cases the notion of time in the quantum
model disappears \cite{isham}. Moreover, even if all these problems are solved,
the question  remains
of the interpretation of the central object of quantum cosmology, the wave function of the Universe. Among the most popular interpretation
schemes for the wave function of the Universe are the many-worlds \cite{tipler} and
the ontological \cite{holland,nelson} interpretations of quantum mechanics.

\par

The many-worlds interpretation differs markedly from
the Copenhaguen interpretation of quantum mechanics since, in a certain sense, the notion of probability is abandoned:
All possibilities are actually realized and new universes are
continously created by acts of observation   according to the different possible
eigenvalues obtained on measurement of an observable. But, in practice, the evolution of observables
such as  the scale factor is followed by means of the evaluation of
expectation values. On the other hand, just like in the ordinary Copenhagen interpretation, the whole structure of Hilbert space and self-adjoint operators
is kept intact.
In its turn, the ontological interpretation of quantum mechanics
makes use of the notion of trajectory. Thus, to a certain extent, the probabilistic concepts
of the Copenhaguen school are also abandoned. This seems to be particularly
interesting when we are discussing a quantum model for the Universe,
since by definition the Universe encompasses everything, and the probabilistic concepts
cannot be applied to it meaningfully.

\par

In a concrete study of a quantum cosmological model, a description for the matter content
must be introduced. In principle, due to the quantum character of the problem, the matter
content should be described by fundamental fields, as done in \cite{kiefer}, for example.
Predictions for the behaviour of the quantum Universe may be obtained, for example, by using
the WKB approximation, establishing trajectories in phase space. However, general
exact solutions are hard to find (even in the minisuperspace), the Hilbert space structure is obscure and it is a subtle matter to
recover the notion of a semiclassical time \cite{isham,kiefer}.

\par
In the present work, we describe matter as a perfect fluid. This description is essentially
semiclassical from the start, but it has the advantage of furnishing a variable, connected
with the matter degrees of freedom, which can naturally be identified with time, leading to a well-defined Hilbert space structure.
Another attractive feature of the phenomenological description is that it allows us to treat
the barotropic equation of state $p = \alpha\rho$ with arbitrary $\alpha$: General exact solutions
can be obtained, constituting a nice ''laboratory" for quantum cosmological models.

\par
To construct a quantum perfect fluid model,
it is very convenient to use  Schutz's formalism \cite{schutz1,schutz2} for a perfect fluid interacting with the gravitational field.
In this formalism, certain velocity potentials are introduced giving dynamics to
the fluid degrees of freedom. After some canonical transformations, at least one
conjugate momentum associated to matter appears linearly in the
action integral, and in this way a Schr\"odinger equation can be obtained
with the matter variable playing the role of time. Therefore, all aparatus
of ordinary quantum mechanics can, in principle, be employed in order
to obtain predictions regarding the evolution of the Universe.

\par
Up to now, quantum perfect fluid models have been constructed only for
the most common equations of state, in particular those corresponding to dust, radiation and
the vacuum \cite{gotay,rubakov,flavio1,nelson1}.
Predictions on the behaviour of the scale
factor of the Universe have been made with the help  of
the many-worlds  as well as  the de Broglie-Bohm interpretations
of quantum mechanics. For those special  equations of state, universes with
a bounce have been found, with absence of singularity; the classical behaviour is recovered for asymptotically large universes.

\par
Here, we  generalize the previous investigations by studying quantum perfect fluid models for any
barotropic equation of state $p = \alpha\rho$. The Wheeler-DeWitt equation
is solved and wave packets are constructed. Next, using both
the many-worlds and ontological frameworks, the behaviour of the
scale factor is determined. For any value of $\alpha$ smaller than
one a singularity-free bouncing Universe is obtained. Asymptotically,
for large values of time, the classical behaviour is recovered.
Although the results for the scale factor are independent of the interpretation
scheme employed, the use of the ontological one allows us to
verify explicitly that a repulsive quantum force appears as the Universe
approaches the primordial singularity, leading to the bounce. The model predicts an accelerated expansion today if
$- 1/3 > \alpha > - 1$. For $\alpha = 1$ it is doubtful whether the quantum model exists at all,
since  we have been unable obtain finite-norm states
due to divergences in the stationary wave functions.

\par
This paper is organized as follows. In the next section, the quantum cosmological
model with a perfect fluid as the matter content is constructed in Schutz's formalism, and
the Wheeler-DeWitt equation in  minisuperspace is written down. The inner product and boundary conditions are given that insure
self-adjointness of the Hamiltonian operator. In
Section 3 the flat case is considered. Wave packets are constructed and the expectation value for
the scale factor is evaluated, as well as the Bohmian trajectories characteristic of
the de Broglie-Bohm formalism. Sections 4 and 5 are dedicated to a brief discussion of the cases of positive or negative curvature. In Section 6 our conclusions are presented.

\section{The quantum model}

We need the Hamiltonian for a perfect fluid model in the formalism
developed by Schutz.
The starting point is the action for gravity plus perfect fluid,
which in this formalism is written as

\begin{equation}
\label{action}
S = \int_Md^4x\sqrt{-g}\, R + 2\int_{\partial M}d^3x\sqrt{h}\, h_{ab}\, K^{ab}
+ \int_Md^4x\sqrt{-g}\, p \quad ,
\end{equation}
where $K^{ab}$ is the extrinsic curvature, and $h_{ab}$ is the induced
metric over the three-dimensional spatial hypersurface, which is
the boundary $\partial M$ of the four dimensional manifold $M$. Units are chosen such that the
factor $16\pi G$ becomes equal to one.
The first two terms were first obtained in \cite{adm};
the last term of (\ref{action}) represents the matter contribution to
the total action, $p$ being the pressure, which is linked to the
energy density by the equation of state $p = \alpha\rho$. In
Schutz's formalism \cite{schutz1,schutz2}
the fluid's four-velocity is expressed in terms of five potentials $\epsilon$,
$\zeta$, $\beta$, $\theta$ and $S$:
\begin{equation}
U_\nu = \frac{1}{\mu}(\epsilon_{,\nu} + \zeta\beta_{,\nu} +
\theta S_{,\nu})
\end{equation}
where $\mu$ is the specific enthalpy. The variable $S$ is the specific
entropy, while the potentials $\zeta$ and $\beta$ are connected with
rotation and are absent of models of the Friedmann-Robertson-Walker (FRW) type. The variables $\epsilon$ and
$\theta$ have no clear physical meaning.
The four-velocity is subject to the normalization condition
\begin{equation}
U^\nu U_\nu = -1 \quad .
\end{equation}

\par
The FRW metric
\begin{equation}
ds^2 = - N^2dt^2 + a^2(t)\sigma_{ab}dx^adx^b \quad ,
\end{equation}
is now inserted in the action (\ref{action}).
In this expression, $N(t)$ is the lapse function and $\sigma_{ab}$ is the metric on the
constant-curvature spatial section.
Using the constraints for the fluid, and after some thermodynamical considerations,
the final reduced action, where surface terms were discarded,
takes the form
\cite{rubakov}
\begin{equation}
S = \int dt\biggr[-6\frac{\dot a^2a}{N} + 6kNa + N^{-1/\alpha}
a^3\frac{\alpha}{(\alpha + 1)^{1/\alpha + 1}}(\dot\epsilon +
\theta\dot S)^{1/\alpha + 1}\exp\biggr(- \frac{S}{\alpha}\biggl)
\biggl] \quad .
\end{equation}
This reduced action may be further simplified  leading,
by canonical methods \cite{rubakov}, to the
super-Hamiltonian
\begin{equation}
{\cal H} = - \frac{p_a^2}{24a} - 6ka + p_\epsilon^{\alpha + 1} a^{-3\alpha}e^S
\end{equation}
where $p_a = -12{\dot aa}/{N}$ and $p_\epsilon = -\rho_0 U^0 Na^3$,
$\rho_0$ being the rest mass density of the fluid.
The  canonical transformation
\begin{equation}
T = -p_Se^{-S}p_\epsilon^{-(\alpha + 1)} \quad , \quad
p_T = p_\epsilon^{\alpha + 1}e^S \quad , \quad
\bar\epsilon = \epsilon - (\alpha + 1)\frac{p_S}{p_\epsilon} \quad ,
\quad \bar p_\epsilon = p_\epsilon \quad ,
\end{equation}
which generalizes the one used in \cite{rubakov},
takes the super-Hamiltonian to the  final
form
\begin{equation}
{\cal H} = - \frac{p_a^2}{24a} - 6ka + \frac{p_T}{a^{3\alpha}} \,\, ,
\end{equation}
where the momentum
$p_T$ is the only remaining canonical variable associated with matter. It appears linearly
in the super-Hamiltonian. The parameter $k$ defines the curvature of the
spatial section, taking the values $0, 1, - 1$ for a flat, positive-curvature or negative-curvature
Universe, as usual.

\par
Imposing the standard quantization conditions on the canonical momenta and demanding that the super-Hamiltonian operator annihilate
the wave function, we are led to the following Wheeler-DeWitt equation in minisuperspace ($\hbar =1$):
\begin{equation}
\label{sle}
\frac{\partial^2\Psi}{\partial a^2} - 144ka^2\Psi + i24a^{1 - 3\alpha}\frac{\partial\Psi}{\partial t} = 0 \quad .
\end{equation}
In this equation, $t=-T$ corresponds to the time coordinate in
a parametrization such that $N = a^{3\alpha}$, as follows from Hamilton's classical equations of motion \cite{flavio1}. By the way, the classical equation of motion for the scale factor is solved in a unified form for any $\alpha\in[0,1]\,$ in terms of hypergeometric functions in \cite{assad}.

\par
Equation (\ref{sle}) takes the form of a Schr\"odinger equation $i\partial\Psi/\partial t = {\hat H} \Psi$. As discussed in \cite{nivaldo, flavio1}, in order for  the Hamiltonian operator ${\hat H}$ to be self-adjoint  the inner product of any two  wave functions $\Phi$ and $\Psi$ must take the form
\begin{equation}
(\Phi,\Psi) = \int_0^\infty a^{1 - 3\alpha}\Phi^*\Psi da \quad \,\, ,
\end{equation}
and restrictive boundary conditions must be imposed to the wave functions in the domain of ${\hat H}$, the simplest ones being
\begin{equation}
\label{boundary}
\Psi(0,t) = 0 \quad \mbox{or} \quad \frac{\partial\Psi (a,t)}{\partial a}\bigg\vert_{a = 0} = 0 \quad .
\end{equation}
\hspace{0.6cm}The Wheeler-DeWitt equation (\ref{sle}) can  be solved by
separation of variables. Indeed, writing
\begin{equation}
\Psi(a,t) = e^{-iEt}\xi(a)
\label{11}
\end{equation}
there results
\begin{equation}
\label{sle2}
\xi'' - 144ka^2\xi +24Ea^{1 - 3\alpha}\xi= 0 \quad ,
\end{equation}
where the prime means derivative with respect to $a$.

\section{The flat case ($k=0$)}

As will be shown below,  for  $\alpha <1$  one can readily build
wave-packet  solutions to the time-dependent  Wheeler-DeWitt  equation  (\ref{sle}) by superposition
of stationary states.

\subsection{Stationary states}

\hspace{0.6cm}For $k = 0$ the time-independent  Wheeler-DeWitt equation (\ref{sle2}) reduces to

\begin{equation}
\label{de1}
\xi'' + 24Ea^{1 - 3\alpha}\xi = 0 \quad .
\end{equation}
If  $\alpha < 1$
it is possible
to show that the parameter $E$ is positive.
Equation (\ref{de1}) admits a solution under the form of
Bessel functions, leading to the following final expression for
the stationary wave functions:
\begin{equation}
\label{stiff}
\Psi_E = e^{-iEt}\sqrt{a}\biggr[c_1J_{\frac{1}{3(1 - \alpha)}}\biggr(\frac{\sqrt{96E}}{3(1 - \alpha)}a^{\frac{3(1 - \alpha)}{2}}\biggl) + c_2Y_{\frac{1}{3(1 - \alpha)}}\biggr(\frac{\sqrt{96E}}{3(1 - \alpha)}a^{\frac{3(1 - \alpha)}{2}}\biggl)\biggl] \quad .
\end{equation}

The particular cases  $\alpha = 0, 1/3$ and $- 1$ have already
been investigated in \cite{rubakov,gotay,flavio1,nivaldo}.
The above solutions are not valid for $\alpha = 1$. In this special case
equation (\ref{de1}) becomes an Euler's type equation, the
general  solution of which takes the form
\begin{equation}
\label{alpha1}
\Psi_E = e^{-iEt}\sqrt{a}\biggr[c_1a^{\frac{\sqrt{1 - 96E}}{2}} +
c_2a^{-\frac{\sqrt{1 - 96E}}{2}}\biggl] \quad .
\end{equation}

\par
All of the above solutions must obey one of the  boundary conditions (\ref{boundary}).
The first one amounts to imposing $c_2 = 0$, while the second one
implies $c_1 = 0$, except again for $\alpha = 1$. The trouble with the two linearly independent stationary solutions for $\alpha = 1$ is that, being just powers of $a$, their behaviour is irregular either at $a=0$ or at $a=\infty$.

\subsection{Wave packets and the behaviour of the scale factor}

None of the stationary solutions found before has finite norm. Hence, wave packets
must be constructed, by superposing those solutions, in order to obtain
wave  functions capable of describing physical states.
The general structure of these superpositions is
\begin{equation}
\Psi(a,t) = \int_0^\infty A(E)\Psi_E(a,t)dE \quad .
\end{equation}
We specialize the discussion from now on to the case $\alpha < 1$ and
$c_2 = 0$. Nothing of substance would be changed had we  chosen $c_1 = 0$ instead.
Defining $r = \frac{\sqrt{96E}}{3(1 - \alpha)}$, simple
analytical expressions for the wavepacket are found if we choose the function
$A(E)$ to be a quasi-gaussian superposition factor:
\begin{equation}
\Psi(a,t) = \sqrt{a}\int_0^\infty r^{\nu + 1}e^{-\gamma r^2 + i\frac{3}{32}r^{2}
(1 - \alpha)^2t}J_\nu(ra^\frac{3(1 - \alpha)}{2})dr \quad ,
\end{equation}
where $\nu = \frac{1}{3(1 - \alpha)}$ and $\gamma$ is an arbitrary positive constant.
The above integral is known \cite{gradshteyn}, and the wave packet takes  the form
\begin{equation}
\label{wp}
\Psi(a,t) = a\frac{e^{-\frac{a^{3(1 - \alpha)}}{4B}}}{(-2B)^{\frac{4 - 3\alpha}{3(1 - \alpha)}}} \quad ,
\end{equation}
where $B = \gamma - i\frac{3}{32}(1 - \alpha)^2t$. Note that the norm of this wave function is finite only if $\alpha < 1$.

\par
Now, we can verify what these
quantum models predict for the behaviour of the scale factor of the Universe.
In order to do this, we adopt first the many-worlds interpretation, and
calculate the expectation value of the scale factor:
\begin{equation}
<a>(t) = \frac{\int_0^\infty a^{1 - 3\alpha}\Psi(a,t)^*a\Psi(a,t)da}
{\int_0^\infty a^{1 - 3\alpha}\Psi(a,t)^*\Psi(a,t)da} \quad .
\end{equation}
The above integrals  are easily computed, leading to
\begin{equation}
<a>(t) \propto \biggr[\frac{9(1 - \alpha)^4}{(32)^2\gamma^2}t^2 + 1\biggl]^\frac{1}{3(1 - \alpha)} \quad .
\end{equation}
These solutions represent, for $\alpha < 1$, a bouncing Universe, with no singularity,
which goes asymptotically to the corresponding flat classical model,
obtained from (\ref{action}),
when $t \rightarrow \infty$:
\begin{equation}
a(t) \propto t^{2/3(1 - \alpha)} \quad .
\end{equation}
In order to fit  observational evidence \cite{perlmutter,riess} that the expansion of the universe is accelerating, one must require
$- \frac{1}{3}> \alpha > - 1$.

\par
It is believed that the results obtained previously by means
of the many-worlds interpretation scheme  coincides with
those that can be obtained using the ontological interpretation of
quantum mechanics \cite{holland,nelson}, in spite of a recent controversy on this issue \cite{neumaier,marchildon,ghose}. In the ontological
interpretation the wave function is written as

\begin{equation}
\Psi = R\, e^{iS}
\end{equation}
where $R$ and $S$ are real functions.
Inserting this expression in the Wheeler-DeWitt  equation (\ref{sle}),
there results

\begin{eqnarray}
\label{hje}
\frac{\partial S}{\partial t} + \frac{1}{24a^{1- 3\alpha}}\biggr(\frac{\partial S}{\partial a}\biggl)^2 + \, Q &=& 0 \quad ,\\
\frac{\partial R}{\partial t} + \frac{1}{12a^{1 - 3\alpha}}\frac{\partial R}{\partial a}\frac{\partial S}{\partial a} +
\frac{1}{24a^{1 - 3\alpha}}R
\frac{\partial^2S}{\partial a^2} &=& 0 \quad ,
\end{eqnarray}
where $Q = - \frac{1}{24a^{1-3\alpha}}\frac{1}{R}\frac{\partial^2R}{\partial a^2}$ is the quantum potential which corrects the
Hamilton-Jacobi equation (\ref{hje}).
When the quantum potential is more important than the classical potential,
we can expect a behaviour deviating from the classical one. Notice that
in the present case the classical potential is zero, since $k = 0$.

\par
The wave function (\ref{wp}) implies
\begin{eqnarray}
R &=& \biggr[4\gamma^2 + \biggr(\frac{3}{16}\biggl)^2(1 - \alpha)^4t^2\biggl]^{-\frac{4 - 3\alpha}{6(1 - \alpha)}}\, a\,
\exp\biggr\{-\frac{\gamma a^{3(1 - \alpha)}}{4\biggr[\gamma^2 + \biggr(\frac{3}{32}\biggl)^2(1 - \alpha)^4t^2\biggl]}\biggl\} \quad , \\
S &=& - \frac{3}{128}\frac{(1 - \alpha)^2a^{3(1 - \alpha)}t}{
\biggr[\gamma^2 + \biggr(\frac{3}{32}\biggl)^2(1 - \alpha)^4t^2\biggl]} +
\frac{(4 - 3\alpha)}{3(1 - \alpha)}\arctan\biggr[\frac{3}{32}\frac{(1 - \alpha)^2t}{\gamma}\biggl] \quad .
\end{eqnarray}
The Bohmian trajectories, which determine the behaviour of the scale factor,
are given by

\begin{equation}
p_a = \frac{\partial S}{\partial a} \quad .
\end{equation}
Using the definition of $p_a$, taking the lapse function
as $N = a^{3\alpha}$, the equation for the Bohmian trajectories becomes
\begin{equation}
512\frac{\dot a}{a} = 3(1 - \alpha)^3\frac{t}{\biggr[\gamma^2 + \biggr(\frac{3}{32}\biggl)^2(1 - \alpha)^4t^2\biggl]}
\end{equation}

which can be easily integrated to
\begin{equation}
\label{bt}
a(t) = a_0 \biggr[\gamma^2 + \biggr(\frac{3}{32}\biggl)^2(1 - \alpha)^4t^2\biggl]^\frac{1}{3(1 - \alpha)} \quad ,
\end{equation}
$a_0$ being an integration constant.
This is essentially the same behaviour found by computation
of the expectation value of the scale factor. The case $\alpha > 1$ is classically forbidden since it predicts a speed of sound greater than the speed of light, and is also forbidden at the quantum level because the norm of the wave function becomes infinite and, strictly speaking, both the de Broglie-Bohm and many-worlds interpretations can not be applied. It is curious, however, that formal Bohmian trajectories do exist for $\alpha > 1$ and suggest a Universe that begins and ends in a singularity.

\par
The quantum potential takes the form
\begin{eqnarray}
Q(a,t) = -
\frac{\gamma}{32}\frac{1 - \alpha}{\biggr[\gamma^2 + \biggr(\frac{3}{32}\biggl)^2(1 - \alpha)^4t^2\biggl]^2}
\biggr\{3\gamma(1 - \alpha)a^{3(1 - \alpha)} - \nonumber\\
(4 - 3\alpha)\biggr[\gamma^2 + \biggr(\frac{3}{32}\biggl)^2(1 - \alpha)^4t^2\biggl]\biggl\} \quad .
\end{eqnarray}
Inserting in this expression the trajectory (\ref{bt}) the  quantum potential can be written in terms of the
scale factor only:

\begin{equation}
Q(a) = \gamma\frac{1 - \alpha}{32}a_0^{3(1 - \alpha)}\frac{(4 - 3\alpha) -
3\gamma(1 - \alpha)a_0^{3(1 - \alpha)}}{a^{3(1 - \alpha)}} \quad .
\end{equation}
From this expression, it is plain to see that for $\alpha < 1$ the quantum effects
become important near the bounce, while they become negligible for
large values of $a$. Hence, asymptotically the scale factor behaves
classically. The force due to the quantum potential
$F_a = - \partial Q(a,t)/\partial a$ is repulsive, leading to the
avoidance of the singularity.

\par
The bad behaviour of the stiff-matter stationary solutions (\ref{alpha1}) either at $a=0$ or at $a=\infty$ has prevented us from finding finite-norm states by superposing them. This leads us to suspect that no perfect fluid quantum cosmological model exists for $\alpha =1$ and $k=0$.

\section{The positive curvature case ($k=1$)}

In this case, the quantum dynamics is  governed by  the Wheeler-DeWitt equation
(\ref{sle}) with $k=1$. We have been unable to find stationary solutions
for arbitrary $\alpha$, therefore we discuss separately the cases for which solutions could be found in terms of known functions. The case of radiation ($\alpha = 1/3$) is omitted since it has already been treated in \cite{nivaldo}.

\subsection{Cosmic strings ($\alpha = -1/3$)}

\par
Inspection of the Wheeler-DeWitt equation (\ref{sle}) shows that although the geometry is closed the quantum dynamics is equivalent to that of the flat model. The stationary solutions take the form

\begin{equation}
\Psi (a,t)=e^{-iEt}\sqrt{a}\left\{C_{1}\, J_{1/4}(\sqrt{-36+6E}\, a^2)+ C_{2}Y_{1/4}(\sqrt{-36+6E}\, a^2)\right\}.
\end{equation}
A wave packet very similar to (\ref{wp}) with $\alpha = -1/3$ can be constructed, and the behaviour  of the scale factor follows the
 pattern of the flat case.

\subsection{Dust ($\alpha = 0$)}

\par
In this case the time-independent Wheeler-DeWitt equation (\ref{sle2}) reduces to

\begin{equation}
-{\xi}^{\prime \prime}(a) + \left(- 24Ea + 144a^{2}\right){\xi}(a)=0\, .
\label{dust1}
\end{equation}
In terms of the new variable $x=12a - E$ we find
\begin{equation}
-\frac{d^{2}\xi}{dx^{2}}+\left[-
\frac{E^{2}}{144}+\frac{x^{2}}{144}
\right]\xi(a) =0.
\label{dust2}
\end{equation}
Equation (\ref{dust2}) is formally identical to the time-independent Schr\"odinger equation for a harmonic oscillator with unit mass and energy $\lambda$:
\begin{equation}
-\frac{d^{2}\xi}{dx^{2}}+\left[- 2\lambda+w^{2}x^{2}\right]\xi(a)
=0, \label{dust3}
\end{equation}

\noindent
where $2\lambda = E^{2}/144$ and $w=1/12$. Inasmuch as the allowed values of $\lambda$ are
$n+1/2$, the possible values of $E$ are
\begin{equation}
E_{n}=\sqrt{12(2n+1)}\,\, , \mbox{\hspace{0.8cm}} n=0,1,2,...\quad .
\label{dust5}
\end{equation}
Thus the  stationary solutions are
\begin{equation}
{\Psi}_{n}(a,t)=e^{-iE_{n}t}{\varphi}_{n}\left(12a - E_{n}\right),
\label{dust6}
\end{equation}

\noindent
where

\begin{equation}
{\varphi}_{n}(x)=H_n\bigg(\frac{x}{\sqrt{12}}\bigg)e^{-x^2/24}\,\, ,
\label{dust7}
\end{equation}
with $H_n$  the $n$-th Hermite polynomial.

\par
The wave functions (\ref{dust6}) look like stationary quantum wormholes as defined by Hawking and Page \cite{hawking}. However, neither of the boundary conditions (\ref{boundary}) can be satisfied by the
wormhole-like wave functions (\ref{dust6}). Thus, at least in the dust case, our perfect fluid model does not support static quantum wormholes, which are ruled out by the requirement that the Hamiltonian operator be self-adjoint.

\subsection{Stiff matter ($\alpha = 1$)}

\par
The general stationary solutions turn out to take the form

\begin{equation} 
\Psi_E (a,t)=e^{-iEt}\,\sqrt{a}\,\left\{
C_{1}\, K_{\nu}(6a^2) + C_{2}\, I_{\nu}(6a^2) \right\}\,\, ,
\end{equation}
where $K_{\nu}$ and $I_{\nu}$ are modified Bessel fuynctions and $\nu = \sqrt{1 - 96 E}/4$. Since $I_{\nu}$ grows exponentially as $a\to\infty$,
here we must set $C_{2}=0$ and as a consequence the first of the boundary conditions (\ref{boundary}) is satisfied. Unfortunately,
however, we have been unable to find explicit finite-norm solutions to the Wheeler-DeWitt equation by superposing stationary states
because integrals over the order of modified Bessel functions are very hard to perform.

\section{The negative curvature case ($k=-1$)}

We managed to find stationary solutions to the Wheeler-DeWitt equation
(\ref{sle}) with $k=-1$ for the same values of $\alpha$ as in the positive-curvature case.

\subsection{Cosmic strings ($\alpha = -1/3$)}

\par
As in the positive-curvature model, in the present case although the geometry is open the quantum dynamics is equivalent to that of the flat model. A wave packet resembling (\ref{wp}) with $\alpha = -1/3$ can be readily constructed,  the behaviour  of the scale factor  following the pattern of the flat case.

\subsection{Dust ($\alpha = 0$)}

The stationary solutions are given by 

\begin{equation}
\label{whittaker}
\Psi (a,t)=e^{-iEt}(12a+E)^{-1/2}\left\{
C_{1}M_{\frac{iE^2}{48},\frac{1}{4}}(\frac{i(12a+E)^2}{12}) +  C_{2}W_{\frac{iE^2}{48},\frac{1}{4}}(\frac{i(12a+E)^2}{12})\right\}
\end{equation}
where $M_{\kappa , \lambda}$ and
$W_{\kappa , \lambda}$ are Whittaker functions, which are related to confluent hypergeometric functions \cite{bateman}. The Whittaker functions in Eq.(\ref{whittaker})  do not automatically vanish at
$a=0$. Thus, in order to satisfy $\Psi(0,t)=0$ it is necessary to take both $C_{1}\neq 0$ and $C_{2}\neq 0$, the same applying to the second of the boundary conditions (\ref{boundary}). The difficulty in dealing with integrals over the order of Whittaker functions has prevented us from obtaining explicit wave packets.

\subsection{Stiff matter ($\alpha = 1$)}

\par
The general stationary solutions turn out to take the form

\begin{equation}
\nonumber
\Psi (a,t)=e^{-iEt}\sqrt{a}\left\{C_{1}J_{\lambda}(6a^2)+C_{2}Y_{\lambda}(6a^2)\right\}
\end{equation}
where $\lambda = \sqrt{1 + 96 E}/4$.  If $\, 0<E<1/96\,$ any of the two boundary conditions (\ref{boundary}) can be implemented, but explicit wave packets could not be found by superposition  of stationary states because very few results are known for integrals over the order of Bessel functions.

\section{Conclusions}

In this work we have investigated minisuperspace FRW quantum cosmological models  with
perfect fluid for any value of the barotropic parameter $\alpha$. In the case of flat spacelike sections
it has been shown that the models are completly solvable, except for
the stiff matter case ($\alpha = 1$), for which no quantum model could
be constructed due to the divergent nature of the stationary wave functions either for large or small scale factor.
The use of  Schutz's formalism for perfect fluids allowed us to
obtain a Schr\"odinger-like Wheeler-DeWitt equation  in which the only remaining  matter degree of freedom
plays the role of time.

\par
Superposing  stationary wave functions, physically acceptable wave packets were constructed in the case of flat spacelike sections. The time evolution of the scale factor
has been determined in two different ways: Evaluating its expectation
value, in the spirit of the many-worlds interpretation
of quantum cosmology, and also computing the Bohmian trajectories of
the ontological interpretation, in spite of controversies concerning the reality of such trajectories \cite{zeh}. In both cases
the result is essentially the same for $\alpha < 1$: A bouncing
singularity-free Universe is obtained.
The use of the ontological interpretation has allowed us to identify
a quantum potential, from which a quantum force was computed.
It acts repulsively as the Universe approaches the singularity,
leading to avoidance of the singularity. In all cases, the
classical behaviour has been recovered asymptotically. A Universe in accelerated expansion today
requires $- 1/3 > \alpha > - 1$.
Near the bounce the behaviour of the scale factor is the same,
irrespective of the value of $\alpha$.

\par
The extension of the previous analysis to a non-flat spatial section leads to
many technical new challenges. The case $\alpha = 1/3$ has already been extensively studied
in the literature \cite{rubakov,nelson1}, and we did not treat it here. In some other specific cases, the wave equation can be solved.
For $\alpha = 0$, the wave function is expressed in terms of Hermite polynomials ($k =  1$) or
Whittaker functions ($k = - 1$). Unfortunately, it seems unavoidable to perform  numerical
integration in order to obtain the wave packets from which the expectation value for the scale
factor could be evaluated. The case $\alpha = - 1/3$ (a cosmic string fluid) can also be
solved, but it brings nothing new with respect to the flat case, since a cosmic string
fluid mimics a curvature term. For $\alpha = 1$, the same problems concerning the finiteness
of the wave function appear, rendering the construction of the wave packet a very hard (or
even impossible) task.

\par
We do not claim that our approach is superior to the more fundamental one based on quantum fields. In a subject still so far away from empirical tests, and so controversial as regards its physical meaning, we believe that several different approaches should be seriously considered and their consequences pushed as far as possible. This is what we tried to do here as concerns perfect fluid FRW quantum cosmological models. However, in order to construct more realistic models, as regards the earliest stages of the Universe, it is clearly important to consider
fundamental fields that can play crucial roles in that primordial
phase. In particular, it may be interesting to  include a conformal scalar field.
We hope to present such a study in the future.

\vspace{0.5cm}

{\bf Acknowledgements:} We thank CNPq (Brazil) for partial financial support.

\end{document}